\documentclass[10pt,twocolumn,letterpaper]{article}
\usepackage{authblk}  
\usepackage{cite}
\usepackage{amsmath,amssymb,amsfonts}
\usepackage{algorithmic}
\usepackage{graphicx}
\usepackage{textcomp}
\usepackage{placeins}

\usepackage[a4paper, top=22mm, bottom=22mm, left=16mm, right=16mm]{geometry}

\usepackage{booktabs}
\usepackage{multirow}
\usepackage[colorlinks=true, linkcolor=black, citecolor=black, urlcolor=black]{hyperref}
\usepackage{float}

\usepackage{caption}
\usepackage{bm}

\makeatletter
\newcommand\corresp[1]{\textsuperscript{*}}
\makeatother

\begin{document}

\title{Undertrained Image Reconstruction for Realistic Degradation in Blind Image Super-Resolution}

\author[1]{Ru Ito}
\author[2$^\ast$]{Supatta Viriyavisuthisakul}
\author[1]{Kazuhiko Kawamoto}
\author[1,3$^\ast$]{Hiroshi Kera}

\affil[1]{Chiba University}
\affil[2]{King Mongkut's University of Technology Thonburi}
\affil[3]{Zuse Institute Berlin}
\affil[$\ast$]{Corresponding authors: \href{mailto:kera@chiba-u.jp}{kera@chiba-u.jp} \& \href{mailto:supatta.viri@kmutt.ac.th}{supatta.viri@kmutt.ac.th}}

\date{}  

\maketitle
\begin{abstract}

Most super-resolution (SR) models struggle with real-world low-resolution (LR) images.
This issue arises because the degradation characteristics in the synthetic datasets differ from those in real-world LR images. 
Since SR models are trained on pairs of high-resolution (HR) and LR images generated by downsampling, they are optimized for simple degradation. However, real-world LR images contain complex degradation caused by factors such as the imaging process and JPEG compression. Due to these differences in degradation characteristics, most SR models perform poorly on real-world LR images.
This study proposes a dataset generation method using undertrained image reconstruction models. These models have the property of reconstructing low-quality images with diverse degradation from input images. By leveraging this property, this study generates LR images with diverse degradation from HR images to construct the datasets.
Fine-tuning pre-trained SR models on our generated datasets improves noise removal and blur reduction, enhancing performance on real-world LR images. Furthermore, an analysis of the datasets reveals that degradation diversity contributes to performance improvements, whereas color differences between HR and LR images may degrade performance.

\end{abstract}

\section{Introduction}\label{sec:introduction}
High-resolution (HR) images are crucial in various fields. For example, in the medical field, MRI images that clearly capture internal tissues and structures are essential for the early detection of diseases \cite{MRI}. In remote sensing, satellite images that provide detailed geographical information are indispensable for accurate weather forecasting \cite{remote}.
However, obtaining such images requires high-performance cameras and lenses, along with continuous maintenance costs. 
super-resolution (SR), a technique for generating HR images from low-resolution (LR) images, plays a crucial role in cost reduction.

    However, many SR models~\cite{SRCNN, FSRCNN, VDSR, ESPCN, EDSR, SRGAN, ESRGAN, RCAN, SwinIR, HAT} still have room for improvement when they are applied to real-world LR images. This issue arises because the degradation characteristics in the synthetic datasets differ significantly from those in real-world LR images, affecting the training process. These models learn to minimize the error between SR images and HR images using HR-LR image pairs. Ideally, these pairs should be collected using imaging equipment, but the high cost of devices and maintenance makes this challenging.
    To avoid this issue, many studies~\cite{SRCNN, FSRCNN, VDSR, ESPCN, EDSR, SRGAN, ESRGAN, RCAN, SwinIR, HAT} generate LR images by downsampling HR images using methods such as bicubic interpolation. Therefore, most SR models are optimized to enhance LR images that contain simple degradations. 
    In contrast, real-world LR images contain complex degradations, including noise, blur, and compression artifacts caused by imaging and JPEG compression. Due to these differences in degradation characteristics, most SR models exhibit reduced performance on real-world LR images.

    To tackle this challenge, referred to as blind image super-resolution, several studies~\cite{RealSR, DRealSR, SR-RAW, City100, SupER, RealSR2, KA, KMSR, DegradationGAN, FSSR, FSSRGAN} explored constructing HR-LR training pairs. One of the techniques is collecting HR-LR image pairs through imaging devices. Some studies, such as \cite{RealSR, DRealSR, SR-RAW, City100, SupER}, conducted image capture at the cost of expensive equipment and maintenance.
    Meanwhile, another approach involves generating pseudo-real-world LR images by applying degradations extracted from real-world LR images to HR images.  For instance, some studies \cite{RealSR2, KA, KMSR} generated LR images by applying estimated blur and noise from real-world LR images to HR images. Furthermore, other studies \cite{DegradationGAN, FSSR, FSSRGAN} learn the distribution of degradations in real-world LR images using  Generative Adversarial Network (GAN) and generate LR images.
    These methods improved SR performance on real-world LR images but still had limitations. One issue is the diversity of scenes in the dataset. The collection of HR–LR pairs limits the dataset to a maximum of 800 scenes, hindering the development of SR models that are capable of handling diverse real-world scenes.  Generating HR--LR pairs addresses this problem by using various HR images. However, this method applies only a limited set of real-world degradations. As a result, models trained on such datasets fail to adapt to diverse real-world degradations, leading to reduced generalization performance.

\begin{figure}[t]
    \centering
    \includegraphics[width=1.0\linewidth]{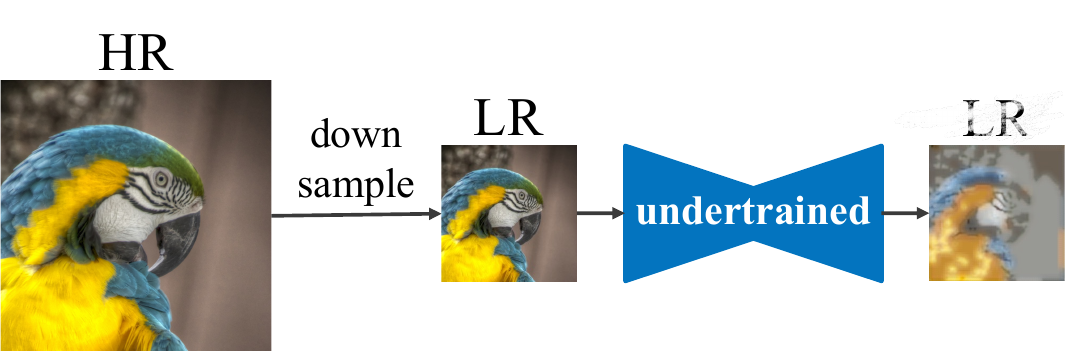}
    \caption{
    The overview of the proposed method. This method first downsamples an HR image to create an LR image. Then, the LR image is fed as an input to the undertrained image reconstruction model to generate a degraded LR image. Finally, it constructs a dataset by pairing the degraded LR image with the HR image.
    }
    \label{fig:method}
\end{figure}

    In this study, we propose a method to generate LR images with diverse degradations from HR images alone to improve SR performance on real-world LR images. The key idea is to utilize the diverse degradations inherent in the outputs of undertrained image reconstruction models.
As illustrated in Figure \ref{fig:method}, the proposed method first downsamples HR images to generate LR counterparts. Then, LR images are processed by an undertrained image reconstruction model, and the resulting outputs are paired with the original HR images.
Fine-tuning pre-trained SR models with this dataset enhances performance on real-world LR images. 
    This method introduces a novel dataset generation approach that utilizes the properties of undertrained image reconstruction models. Additionally, it enables performance improvement through fine-tuning alone, providing a cost-effective and model-agnostic solution.

\begin{figure*}[t]
    \centering
    \includegraphics[width=1.0\linewidth]{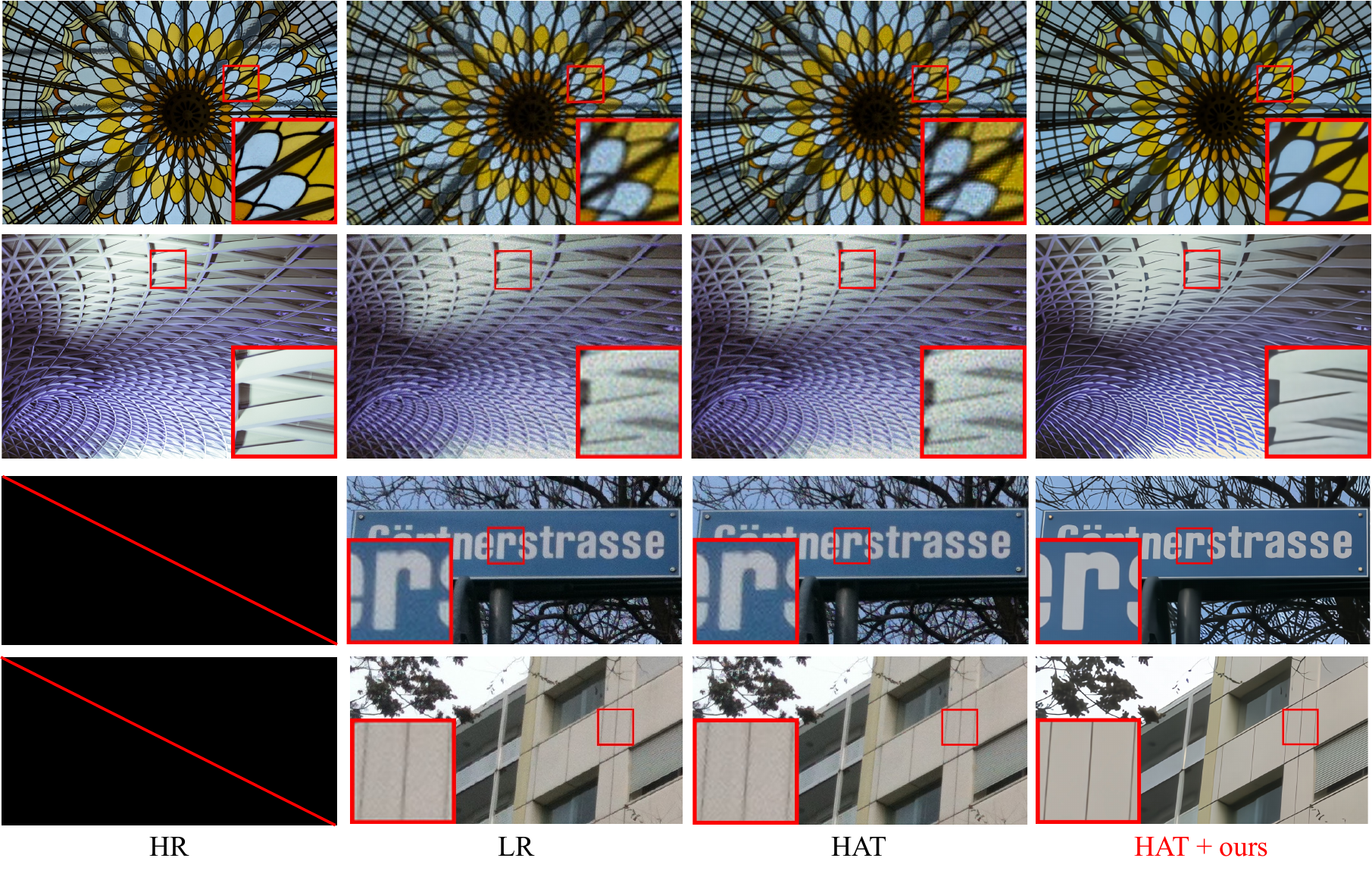}
    \caption{
    Fine-tuning a pre-trained HAT model with a dataset generated by VQ-VAE-2 trained for 8 epochs produced notable improvements. The pre-trained HAT model retained noise and blur from the LR images, but the fine-tuned model effectively reduced noise and restored sharpness.
    }
    \label{fig:top-image}
\end{figure*}

In the experiments, we generated datasets using three image reconstruction models—VAE~\cite{VQVAE}, VQ-VAE-2~\cite{VQVAE2}, and MAE~\cite{MAE}—which were selected through our preliminary experiments. The generated dataset was confirmed to contain various degradations, including noise, ringing, aliasing, and blur.  Fine-tuning was performed on pre-trained SR models, including EDSR~\cite{EDSR}, ESRGAN~\cite{ESRGAN}, SwinIR~\cite{SwinIR}, and HAT~\cite{HAT}. The performance was evaluated on NTIRE2018 Track 3~\cite{NTIRE2018} and NTIRE2020 Track 2~\cite{NTIRE2020}. The results showed that the dataset generated by VQ-VAE-2, particularly with 8 epochs of training, achieved the highest performance.  Specifically, SSIM improved by up to 0.1266, and LPIPS enhanced by down to 0.0816. Additionally, as shown in Figure \ref{fig:top-image}, the dataset contributed to noise removal and blur reduction.
    Further analysis of the generated dataset revealed that the diversity of degradations can improve SR performance. However, among the degradations, color differences between HR and LR images can negatively impact performance.

Our contributions are as follows:
\begin{itemize}
  \item We propose a method that generates LR images with diverse degradations using only HR images by leveraging undertrained image generation models.
  \item Experimental results show that the dataset generated by VQ-VAE-2 improves SR performance on real-world LR images.
  \item It is demonstrated that while degradation diversity in the dataset enhances SR performance, color differences between HR and LR images can negatively impact performance.
\end{itemize}

\section{Related Work} \label{sec:related_work}
In this section, we explored dataset collection using imaging equipment and techniques for generating pseudo-real-world LR images to enhance the performance of SR models on real-world LR images. \\

\noindent \textbf{Dataset collection using imaging equipment.}
One of the earliest datasets, SupER \cite{SupER}, contains over 80,000 pairs of HR-LR images, utilizing the hardware binning technique of a CMOS camera. However, all images are monochrome, limiting its applicability.
    In contrast, City100 \cite{City100} comprises 100 color image pairs captured using two different cameras, a Nikon DSLR and an iPhone. However, this dataset has the limitation of lacking outdoor scenes.  Those were constructed by photographing printed postcards of urban scenes indoors.
    SR-RAW \cite{SR-RAW} and RealSR \cite{RealSR} were proposed to encompass a wider variety of scenes. SR-RAW captures approximately 500 indoor and outdoor scenes using a Sony zoom lens. RealSR includes around 600 indoor and outdoor scenes collected using Nikon and Canon DSLR cameras. While, DRealSR \cite{DRealSR}, an extension of RealSR, expands the dataset further by capturing around 800 scenes with five different DSLR cameras from Sony, Canon, Olympus, Nikon, and Panasonic.
    These datasets provide pairs of real-world HR-LR images. However, those datasets primarily contain static subjects, such as buildings and signs, with few dynamic objects, such as animals. Since the datasets lacked scene diversity, it was difficult for SR models to handle diverse real-world scenes.\\

\noindent \textbf{Generation of pseudo-real-world LR images.} 
    Applying degradation processing to diverse HR images and generating pseudo-real-world LR images can improve scene diversity.
    Several studies have proposed dataset generation, that utilizes real-world LR images. For example, blur kernels and noise estimated from real-world LR images are applied to HR images in \cite{RealSR2} to generate pseudo-real-world LR images.
    While an approach in \cite{KMSR} utilizes a GAN \cite{GAN} to augment estimated blur kernels and produce diverse degradation kernels. Furthermore,  \cite{KA} combining GAN-based kernel augmentation with Stochastic Variation can add natural randomness to generated images.
    A method for generating LR images using a generation model with HR images as input has also been proposed. For example, a GAN generator trained on real-world LR images is used in \cite{DegradationGAN} to generate LR images from HR images. However, GAN training is often unstable, leading to image corruption. To improve training stability, an adversarial loss is applied only to high-frequency components \cite{FSSR}. Additionally, a color attention module is introduced to mitigate color shifts between generated HR and LR images \cite{FSSRGAN}.
    These methods performed well in generating real-world LR images. However, the generated image captures only a subset of real-world degradations, restricting the variety of degradations in the dataset and reducing model adaptability to real-world scenarios.\\

We propose a method to generate LR images with diverse degradations from HR images alone. Similar to previous studies, we aim to create a dataset that improves performance on real-world LR images. To the best of our knowledge, this approach introduces a novel method not explored in existing studies.

\section{Proposed Method and Preliminary Experiments}\label{sec:propose}
This study proposes a dataset generation method that utilizes undertrained image reconstruction models. The generated dataset contains diverse degradations. Then, it is used to train the SR models to improve performance on real-world LR images.

\subsection{Dataset generation}

First, our proposed method downsamples an HR image $\mathbf{y}$ to obtain an LR image $\mathbf{x}$. Then, the LR image $\mathbf{x}$ is input into a network $G_{\theta}:\mathbf{x} \to \mathbf{x} $ parameterized by $\theta$. 
    The key point is that utilizing an undertrained network $G_{\theta}:\mathbf{x} \to \mathbf{x}_{\mathrm{deg}} $ generates a degraded LR image $\mathbf{x}_{\mathrm{deg}}$ with diverse degradations, as shown in Figure~\ref{fig:method}.
    Finally, the dataset $\mathcal{D}$ is constructed by pairing the HR image $\mathbf{y}$ with the degraded LR image $\mathbf{x}_{\mathrm{deg}}$. The dataset is represented as follows:

\begin{align} \label{eq:dataset}
    \mathcal{D} =  \{(\mathbf{x}^i_{\mathrm{deg}}, \mathbf{y}^i)\}^{|\mathcal{D}|}_{i=1}
\end{align}
Here, $i$ represents the index corresponding to each data pair in the dataset $\mathcal{D}$.
Fine-tuning a pre-trained SR model $f_{\theta}$ with this dataset enhances SR performance on real-world LR images.

\subsection{Performance Comparison of Image Reconstruction Models
}\label{sec:compare}
\begin{figure}[t]
    \centering
        \includegraphics[width=1\linewidth]{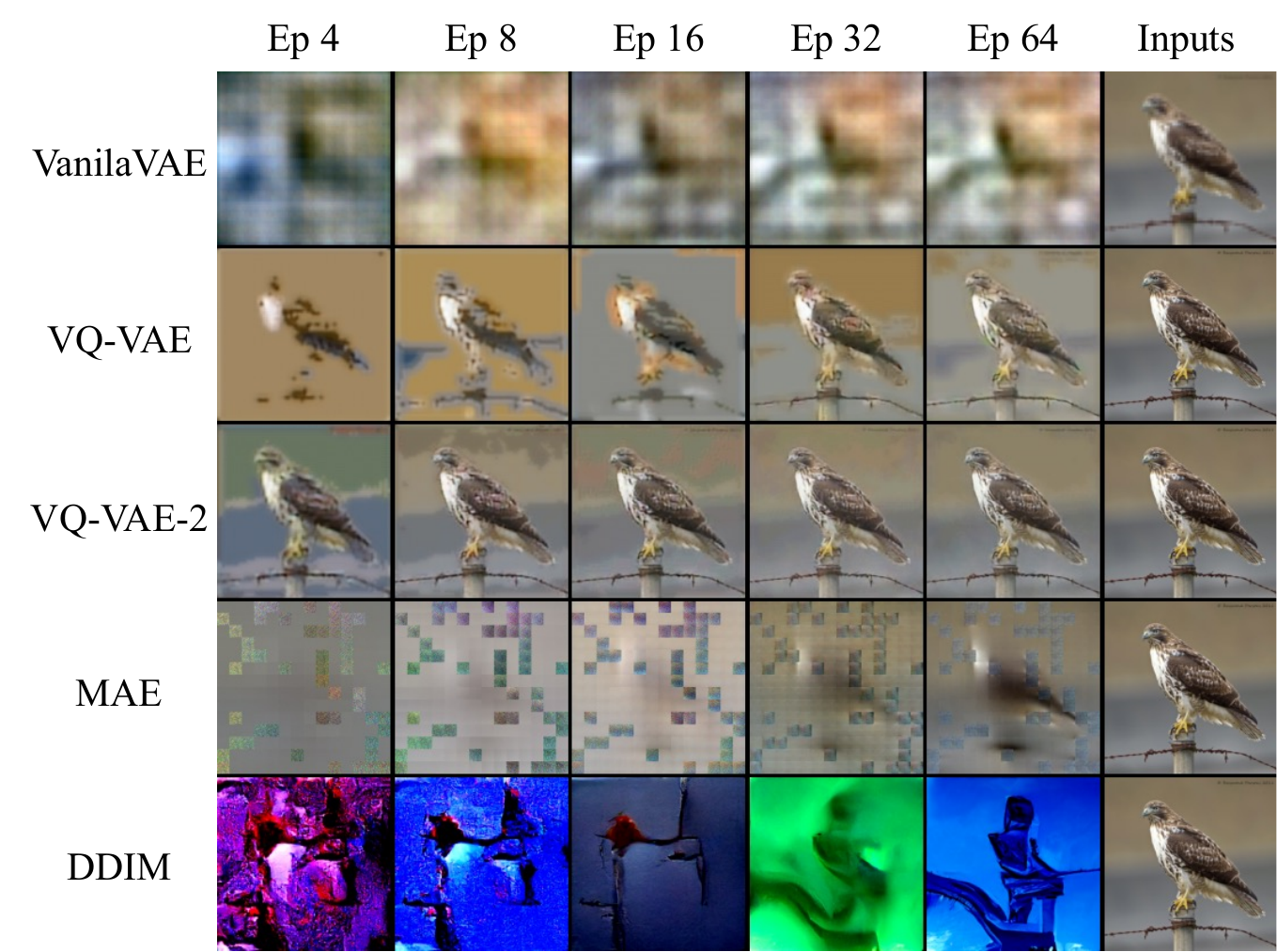}
    \caption{
    Reconstruction results of five image reconstruction models. The rightmost column represents the input images, and the other columns show reconstructed images at different training epochs. The results indicate that Vanilla VAE and DDIM fail to reconstruct the input images. MAE produces images that deviate significantly from the input at epochs 4, 8, and 16, but at epochs 32 and 64, it reconstructs structural information and color to some extent. VQ-VAE and VQ-VAE-2 achieve relatively accurate reconstructions even with minimal training while incorporating degradation.
    }
    \label{fig:taka}
\end{figure}

As a preliminary experiment, we compare the performance of representative image reconstruction models to select a suitable model for this method.
Emphasis was placed on the ability to reconstruct images that not only preserve the structural information of the input but also retain degradations similar to real-world LR images. The following describes the experimental setup and the results. \\

\noindent \textbf{Setup.}
The experiment used five image reconstruction models: VanillaVAE~\cite{VAE}, VQ-VAE~\cite{VQVAE}, VQ-VAE-2~\cite{VQVAE2}, MAE~\cite{MAE}, and DDIM~\cite{DDIM}. These models were trained for 4, 8, 16, 32, and 64 epochs using Tiny-ImageNet, which contained 45,000 images extracted from ImageNet~\cite{ImageNet}. In Figure \ref{fig:taka}, the sample of the result confirmed that the number of training epochs was insufficient for all models.
The input image sizes for each model were set as follows: 64$\times$64 for VanillaVAE, 224$\times$224 for MAE, and 256$\times$256 for VQ-VAE, VQ-VAE-2, and DDIM. MAE was configured to reconstruct the entire image instead of only the masked regions. The number of sampling steps for DDIM was set to 100. GAN-based models~\cite{GAN} were not used because no suitable models exist for image reconstruction. The comparison was conducted through a qualitative evaluation of the reconstructed images. \\

\noindent \textbf{Results.}
We compared the performance of five image reconstruction models and selected VQ-VAE, VQ-VAE-2, and MAE as the models used in the proposed method.
Figure \ref{fig:taka} presents the reconstruction results of each model. The rightmost column shows the input images, and the other columns display reconstructed images at different training epochs. The result demonstrates that reconstructions by VanillaVAE and DDIM are significantly different from the input images. 
    The poor reconstruction quality of VanillaVAE can be attributed to its limited representation capacity. Specifically, VanillaVAE constrains the latent space to a standard normal distribution, which is effective for reconstructing relatively simple structures, such as facial images. However, this simple latent representation is insufficient for complex and diverse datasets such as ImageNet~\cite{ImageNet}, leading to a significant decline in reconstruction performance. 
    For DDIM, the poor reconstruction quality is attributed to instability in noise removal during the reverse process. Without sufficient training, noise prediction in the reverse process becomes unstable, causing cumulative errors at each step. Consequently, the reconstructed images are significantly different from the input.
    MAE produces reconstructions that differ greatly from the input when trained with few epochs. However, at epochs 32 and 64, it captures structural and color information to some extent. Similarly, VQ-VAE and VQ-VAE-2 can reconstruct images reasonably well, even with minimal training and incorporating degradation. 
    Based on these findings, this study employs VQ-VAE, VQ-VAE-2, and MAE to generate datasets, as these models can preserve structural information while introducing degradation.

\section{Experiments}\label{sec:experiments}
This section presents the experimental setup and results of generating datasets using undertrained image reconstruction models and fine-tuning pre-trained SR models. Additionally, we analyze the datasets and discuss their effectiveness.

\subsection{Dataset generation} \label{sec:dataset_generation}
\textbf{Setup.} To improve SR performance on real-world LR images, training datasets are generated using three image reconstruction models: VQ-VAE, VQ-VAE-2, and MAE. These three models are undertrained, following the experimental setup in Section \ref
{sec:compare}.
As a pre-processing step, HR images from the DF2K (DIV2K + Flickr2K) \cite{DF2K} dataset are cropped to 1,024 × 1,024, preparing approximately 20,000 images. These images are then downsampled with a scaling factor of 4.

\begin{figure*}[!t]
    \centering
    \includegraphics[width=0.8\linewidth]{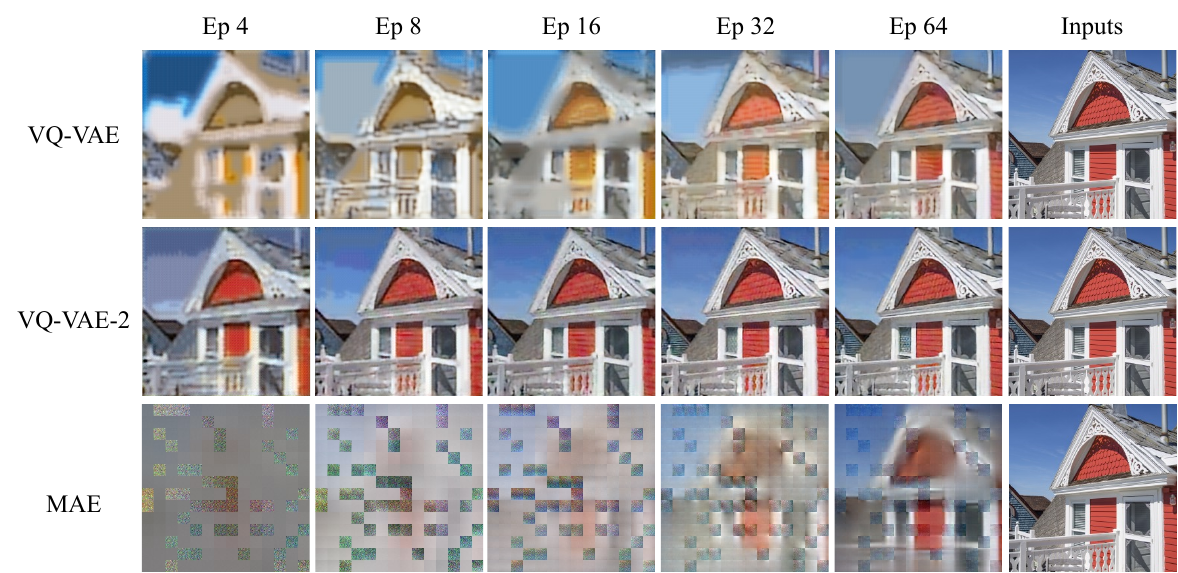}
    \caption{
    Examples of LR images generated by each model. The rightmost column represents the input images, while the other columns show the reconstructed images. All models preserve the structural information of the input while introducing degradation. VQ-VAE-2 reconstructs images close to the input regardless of the number of training epochs. VQ-VAE tends to produce images with a yellowish tint, and MAE introduces sparse degradations. These results confirm that each model generates degraded images with distinct characteristics.
    }
    \label{fig:dataset}
\end{figure*}

Figure \ref{fig:dataset} shows the examples of generated LR images by each model. The rightmost column corresponds to the input images, while the other columns show the reconstructed images from each model. Across all models, the reconstructed images preserve the structural information of the input while incorporating degradation.  
VQ-VAE-2 reconstructs images close to the input regardless of the number of training epochs. VQ-VAE tends to produce images with a yellowish tint, and MAE introduces sparse degradations. These results confirm that each model generates degraded images with distinct characteristics.

\begin{figure*}[!t]
    \centering
    \includegraphics[width=0.9\linewidth]{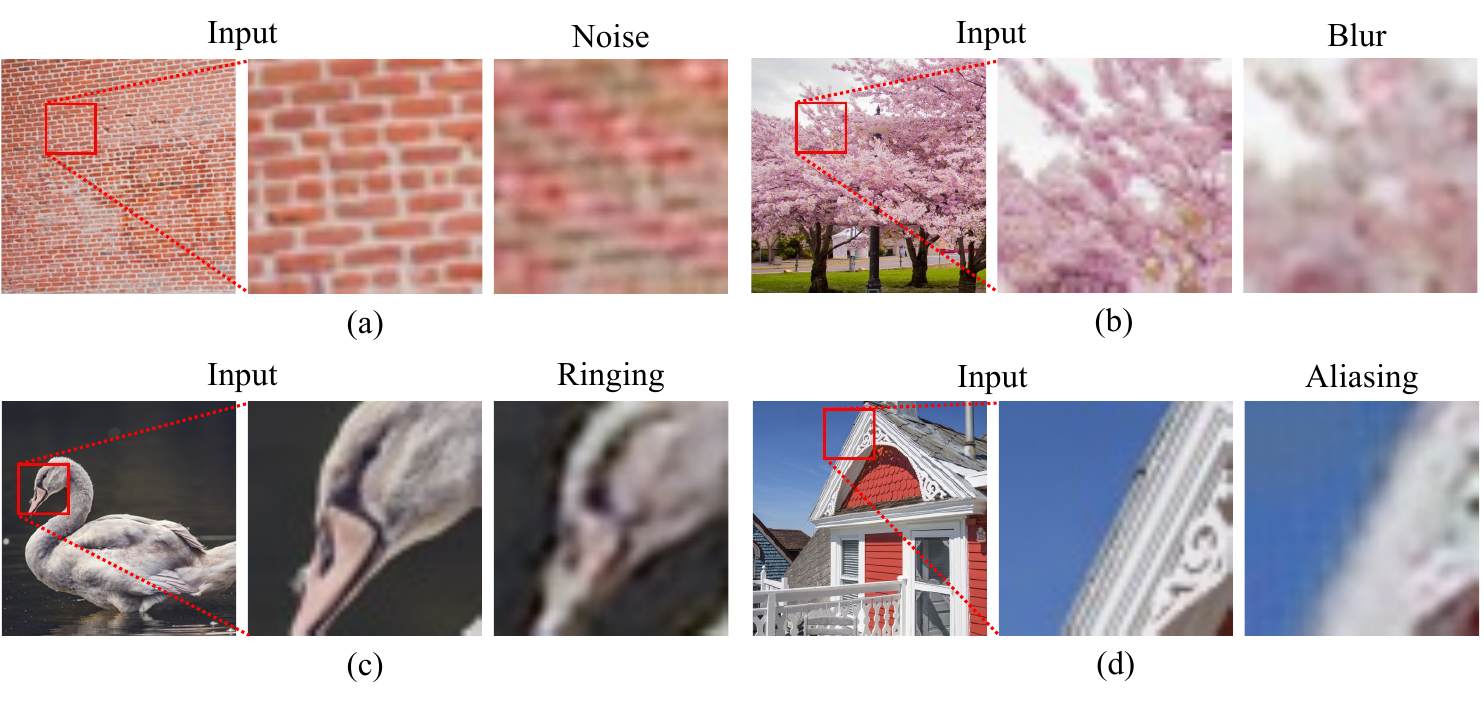}
    \caption{
    Examples from the dataset generated by VQ-VAE-2 trained for epoch 8. The left two columns show input images, and the right represents reconstructed images.  
    In (a), the brick colors vary across red and green, introducing noise. In (b), the cherry blossom contours appear blurred, indicating the presence of blur. In (c), ring-shaped artifacts are visible around the head of the duck, resulting in ringing. In (d), jagged edges appear along the straight lines of the roof, showing aliasing. These results demonstrate that the proposed method can generate datasets containing diverse types of degradation.
    }
    \label{fig:dataset_deg}
\end{figure*}

Furthermore, Figure \ref{fig:dataset_deg} shows examples of LR images from the dataset generated by VQ-VAE-2 trained for epoch 8. The first two columns of each pane show the input image and its magnified view, and the last column displays the reconstructed images with various types of degradation.
In (a), the brick colors vary across red and green, introducing noise. In (b), the cherry blossom contours appear blurred, indicating the presence of blur. In (c), ring-shaped artifacts are visible around the head of the duck, resulting in ringing. In (d), jagged edges appear along the straight lines of the roof, showing aliasing.  
These results demonstrate that the proposed method can generate datasets containing diverse types of degradation.

\subsection{Fine-tuning pre-trained models} \label{sec:fine-tuning}
We performed fine-tuning on the pre-trained SR model, HAT~\cite{HAT}, using the generated datasets to assess its effectiveness in enhancing SR performance on real-world LR images. The results demonstrate the effectiveness of the dataset generated by VQ-VAE-2. To further verify its general applicability, we additionally fine-tuned three pre-trained SR models—EDSR~\cite{EDSR}, ESRGAN~\cite{ESRGAN}, and SwinIR~\cite{SwinIR}—using the same datasets. Performance evaluation was conducted using the NTIRE2018 Track 3~\cite{NTIRE2018} and NTIRE2020 Track 2~\cite{NTIRE2020} datasets.

\subsubsection{Validation of Dataset Effectiveness} \label{sec:hat-finetune}

We do fine-tuning the pre-trained HAT model using datasets generated by VQ-VAE, VQ-VAE-2, and MAE to assess its effectiveness. The quantitative evaluation results on NTIRE2018 Track 3 are shown in Table \ref{tab:finetune-hat}, where the best values in each column are highlighted in red and the second-best in blue.
It can prove that VQ-VAE-2-generated datasets improve SR performance. Specifically, fine-tuning with datasets from training at epochs 4, 8, 32, and 64 resulted in improvements in SSIM and LPIPS. Notably, the dataset from epoch 8 improved SSIM by 0.0951 points and LPIPS by 0.0371 points.

\begin{figure}[t]
    \centering
    \includegraphics[width=1\linewidth]{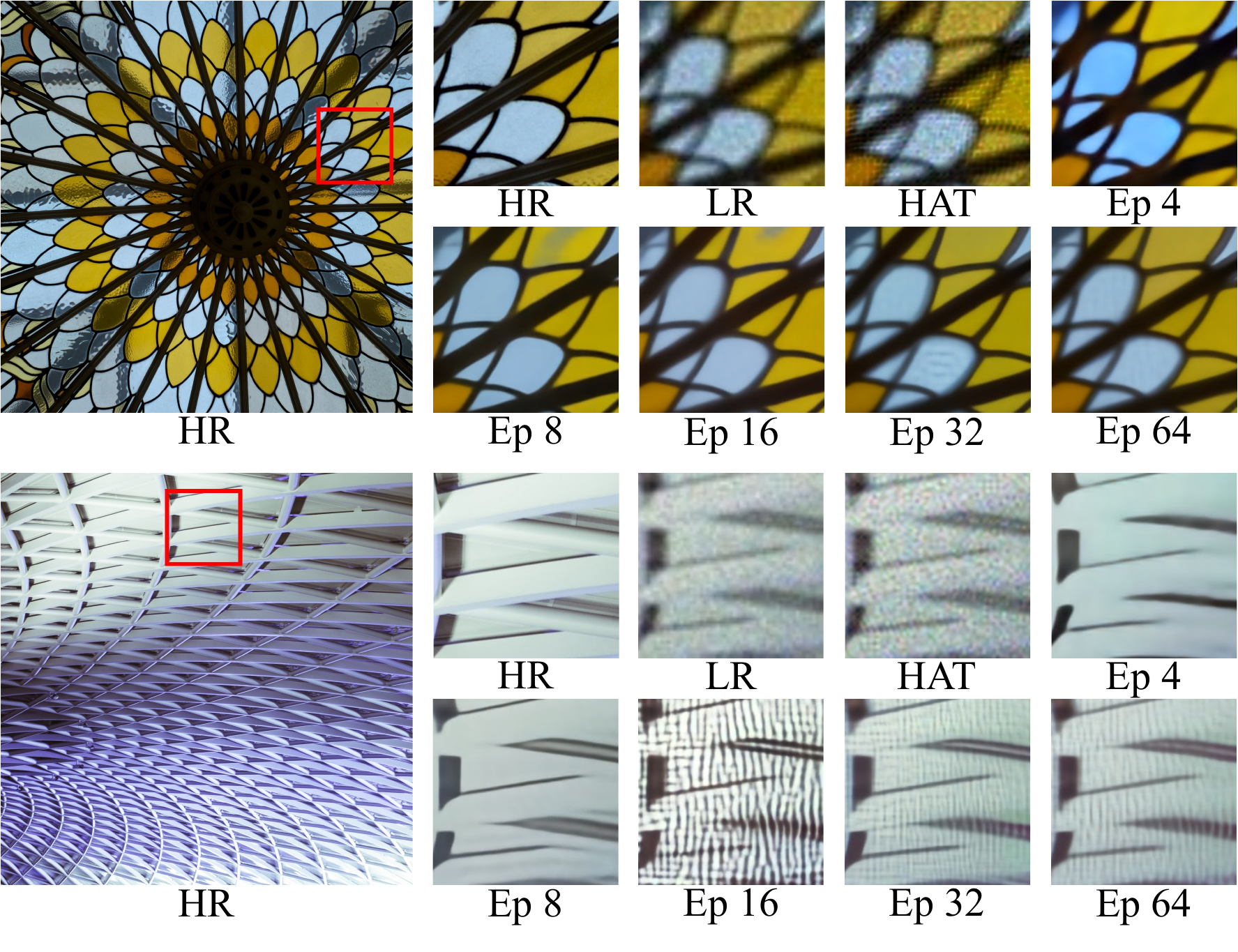}
    \caption{
    The result of fine-tuning the pre-trained HAT with the dataset generated by VQ-VAE-2. The SR images from the pre-trained model retain noise and blur, while fine-tuning improves these artifacts. However, using the dataset from epoch 4 removes noise and blur but tends to produce highly saturated images. Datasets from epochs 16, 32, and 64 introduce noticeable artifacts. In contrast, the dataset from epoch 8 eliminates noise, blur, and artifacts, producing high-quality images close to the HR image.
    }
    \label{fig:vqvae2_hat}
\end{figure}

Figure \ref{fig:vqvae2_hat} shows the qualitative evaluation of models fine-tuned with VQ-VAE-2-generated datasets. This compares HR, LR, and SR images by magnifying the regions highlighted in red. 
    The results show that SR images from the pre-trained model HAT retain noise and blur, while the fine-tuned model reduces these artifacts. In particular, the model trained with the dataset from epoch 8 generates images with fewer artifacts and higher visual quality, closely resembling HR images. However, the model fine-tuned with the dataset from epoch 4 successfully reduces noise and blur yet tends to produce highly saturated images. Models fine-tuned with datasets from epochs 16, 32, and 64 exhibit noticeable artifacts. 

Figure \ref{fig:top-image} shows the qualitative evaluation of the model fine-tuned with the dataset from 8 epochs, which showed the greatest improvement. This figure shows that the pre-trained HAT model retained noise and blur from the LR images, but the fine-tuned model effectively reduced noise and restored sharpness.
These results demonstrate that datasets generated by VQ-VAE-2, particularly that from epoch 8, are effective in improving SR performance on real-world LR images.

\begin{table}[!t]
    \centering
    \footnotesize
    \caption{
    The result of fine-tuning the pre-trained HAT with the proposed datasets. The best values in each column are highlighted in red, and the second-best in blue. These results demonstrate the effectiveness of datasets generated by VQ-VAE-2. Specifically, fine-tuning with datasets from epochs 4, 8, 32, and 64 of VQ-VAE-2 training improved SSIM and LPIPS. Notably, the dataset from epoch 8 improved SSIM by 0.0951 points and LPIPS by 0.0371 points.
    }
    \label{tab:finetune-hat}
    \resizebox{\columnwidth}{!}{ 
        \begin{tabular}{l|l||ccc}
            \toprule
            
            \multirow{2}{*}{Generative Model} & 
            \multirow{2}{*}{Dataset} & 
            \multicolumn{3}{c}{HAT} \\
            
            ~ & ~ & PSNR$\uparrow$ & SSIM$\uparrow$ & LPIPS$\downarrow$ \\
            \midrule
            ~     & Ep 4  & 12.92& 0.4394 & 0.7453 \\
            ~     & Ep 8  & 12.68& 0.4067 & 0.7296 \\
            VQ-VAE & Ep 16 & 12.91& 0.4038 & 0.6891 \\
            ~     & Ep 32 & 11.48& 0.4151 & 0.6855 \\
            ~     & Ep 64 & 9.345& 0.2886 & 0.6753 \\
            \midrule
            ~     & Ep 4  & 16.68& \textcolor{blue}{0.5077} & 0.6386 \\
            ~     & Ep 8  & 18.10& \textcolor{red}{0.5288} & \textcolor{blue}{0.5977} \\
            VQ-VAE-2& Ep 16 & 15.83& 0.2848 & 0.6410 \\
            ~     & Ep 32 & 17.92& 0.4435 & 0.6036 \\
            ~     & Ep 64 & \textcolor{blue}{18.13}& 0.4968 & \textcolor{red}{0.5901} \\
            \midrule
            ~     & Ep 4  & 10.99& 0.2742 & 0.7164 \\
            ~     & Ep 8  & 11.38& 0.3007 & 0.7226 \\
            MAE   & Ep 16 & 11.04& 0.2471 & 0.7133 \\
            ~     & Ep 32 & 11.53& 0.2625 & 0.6904 \\
            ~     & Ep 64 & 15.72& 0.4404 & 0.6348 \\
            \midrule
            \multicolumn{2}{c||}{w/o fine-tune} & \textcolor{red}{18.24}& 0.4337 & 0.6272 \\
            \bottomrule
        \end{tabular}
    }
\end{table}

\subsubsection{Validation of Dataset Generalization} \label{sec:generalization}
\begin{figure*}[h]
    \centering
    \includegraphics[width=0.9\linewidth]{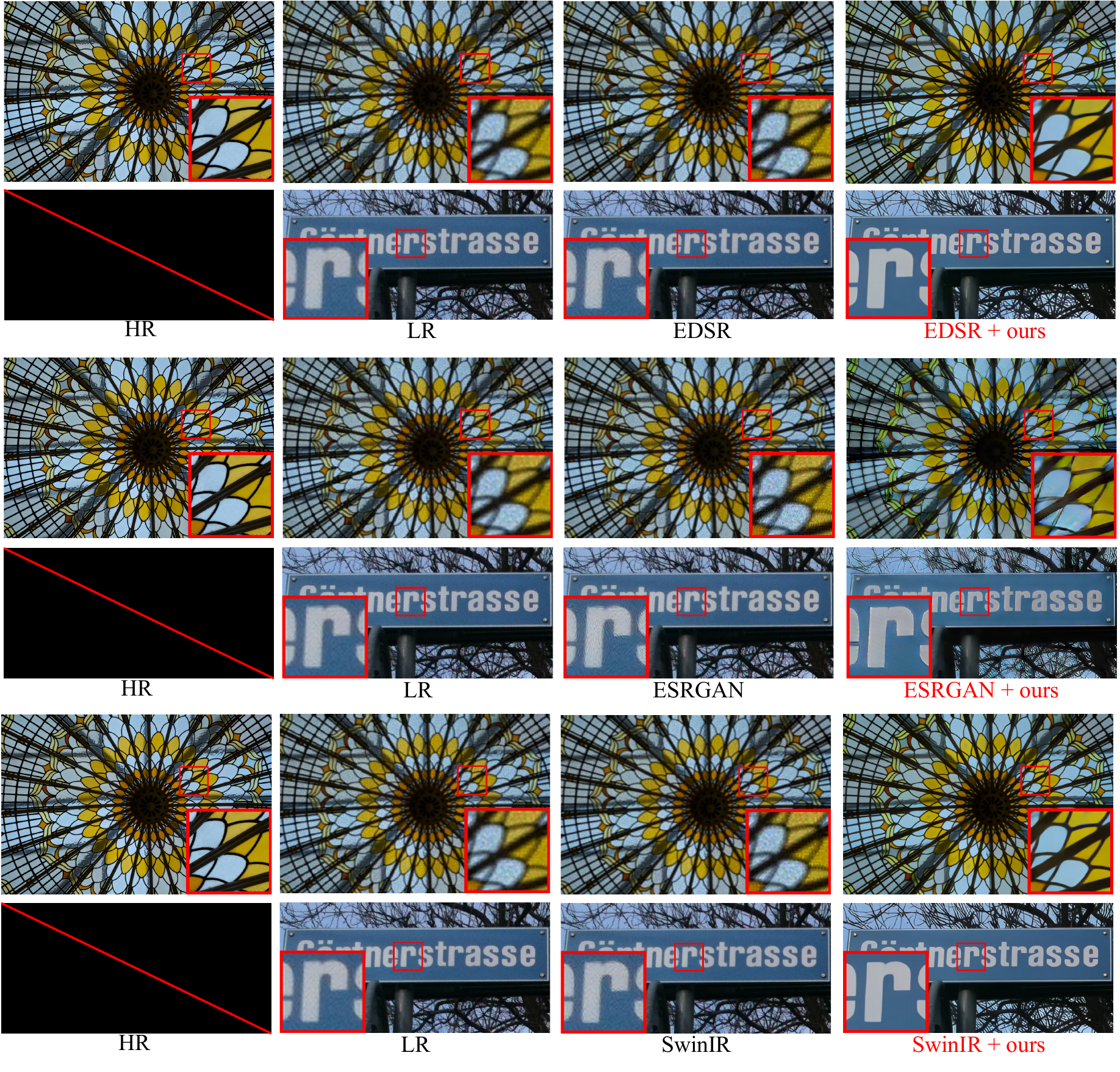}
    \caption{
    The results of fine-tuning EDSR, ESRGAN, and SwinIR with the dataset generated by VQ-VAE-2 trained for epoch 8. While SR images from the pre-trained models retain residual noise and blurring, the fine-tuned models show noticeable improvements in reducing these artifacts.
    }
    \label{fig:edsr_esrgan_swinir}
\end{figure*}

The performance of SR models on real-world LR images is improved by datasets generated by VQ-VAE-2. To demonstrate that the dataset is model-independent, we fine-tuned EDSR~\cite{EDSR}, ESRGAN~\cite{ESRGAN}, and SwinIR~\cite{SwinIR}.
Table~\ref{tab:finetune-models} shows the quantitative evaluation using NTIRE2018 Track 3. The highest values in each column are highlighted in red, and the second-highest values are in blue. 
The table shows that VQ-VAE-2-generated datasets improve SSIM and LPIPS across all three models, confirming their general applicability. In particular, when fine-tuning with the dataset from epoch 8, EDSR improves SSIM by 0.0871 points and LPIPS by 0.0138 points. ESRGAN shows improvements of 0.1266 points in SSIM and 0.0816 points in LPIPS, and SwinIR achieves a 0.0503 points increase in SSIM.
Figure~\ref{fig:edsr_esrgan_swinir} presents the qualitative evaluation of fine-tuned models with the dataset from epoch 8. SR images generated by the pre-trained EDSR, ESRGAN, and SwinIR models exhibit residual noise and blurring. In contrast, fine-tuning models with the dataset reduces noise and blurring, leading to higher-quality SR images.  
These results demonstrate that the effectiveness of the VQ-VAE-2-generated datasets is independent of the SR models.

\begin{table*}[t]
    \centering
    \footnotesize
    \caption{
    The result of fine-tuning the pre-trained EDSR, ESRGAN, and SwinIR with the VQ-VAE-2-generated dataset. Red indicates the highest value in each column, and blue represents the second highest. SSIM and LPIPS improved across all three models. In particular, when fine-tuning with the dataset from epoch 8, EDSR improves SSIM by 0.0871 points and LPIPS by 0.0138 points. ESRGAN shows improvements of 0.1266 points in SSIM and 0.0816 points in LPIPS, and SwinIR achieves a 0.0503 points increase in SSIM.
    }
    \label{tab:finetune-models}
    \resizebox{\textwidth}{!}{ 
    \begin{tabular}{l|l||ccc|ccc|ccc}
        \toprule
        
        \multirow{2}{*}{Generative Model} & 
        \multirow{2}{*}{Dataset} & 
        \multicolumn{3}{c|}{EDSR} &
        \multicolumn{3}{c|}{ESRGAN} &
        \multicolumn{3}{c}{SwinIR} \\ 

        ~ & ~ & PSNR$\uparrow$ & SSIM$\uparrow$ & LPIPS$\downarrow$ & 
                PSNR$\uparrow$ & SSIM$\uparrow$ & LPIPS$\downarrow$ & 
                PSNR$\uparrow$ & SSIM$\uparrow$ & LPIPS$\downarrow$ \\
        \midrule
        
        ~     & Ep 4  & 16.35& \textcolor{blue}{0.5034} & 0.6691
                        & 14.80& 0.2715 & 0.6513 
                        & 16.15& \textcolor{red}{0.4903} & 0.6496 \\
        ~     & Ep 8  & \textcolor{blue}{18.16}& \textcolor{red}{0.5238} & \textcolor{red}{0.6123} 
                        & \textcolor{blue}{16.60}& \textcolor{red}{0.3981} & \textcolor{red}{0.5814} 
                        & 17.70& \textcolor{blue}{0.4842} & 0.6169 \\
        VQ-VAE-2& Ep 16 & 16.66& 0.3599 & 0.6327 
                        & 16.07& 0.3765 & 0.6007 
                        & 15.64& 0.2835 & 0.6409 \\
        ~     & Ep 32 & 17.15& 0.3786 & \textcolor{blue}{0.6131} 
                        & 15.50& \textcolor{blue}{0.3918} & \textcolor{blue}{0.5932} 
                        & 16.94& 0.4187 & \textcolor{blue}{0.6089} \\
        ~     & Ep 64 & 17.55& 0.384 & 0.614 
                        & 15.02& 0.2697 & 0.6424 
                        & \textcolor{blue}{17.90}& 0.4392 & \textcolor{red}{0.6051} \\
        
        \midrule
        
        \multicolumn{2}{c||}{w/o fine-tune} 
        & \textcolor{red}{18.23}& 0.4367 & 0.6261
        & \textcolor{red}{17.64}& 0.2715 & 0.663
        & \textcolor{red}{18.23}& 0.4339 & 0.6275 \\
        
        \bottomrule
    \end{tabular}}
\end{table*}

\subsection{Analysis of datasets} \label{sec:dataset_analysis}

This section analyzes the dataset generated by the proposed method and discusses the factors that contributed to its effectiveness or lack of effectiveness.

\subsubsection{Degradation Diversity} \label{sec:deg}
We quantitatively evaluate the diversity of degradations in each of our datasets and discuss its relationship with effectiveness.
This evaluation is based on the idea that constructing an SR model capable of handling various degradations in real-world LR images requires a training dataset with similar degradation diversity. Degradation diversity is defined by the following entropy $H$ where a higher value indicates greater diversity:

\begin{align} 
    H = - \sum_{\mathbf{x} \in \mathcal{X}} P(\mathbf{x}) \log P(\mathbf{x})
\end{align}

Here, $\mathcal{X}$ represents the set of image degradation classes. $P(\cdot)$ denotes the proportion of each degradation class in the dataset and is obtained using an image classification model capable of categorizing degradations.

Here, we used a ResNet-152~\cite{ResNet-152} pre-trained model on the Tiny-ImageNet-C dataset, which contains 15 types of degradations, including Gaussian blur, Gaussian noise, and JPEG compression. Each degradation class consists of 2,000 images.

\begin{figure}[!t]
    \centering    \includegraphics[width=1\linewidth]{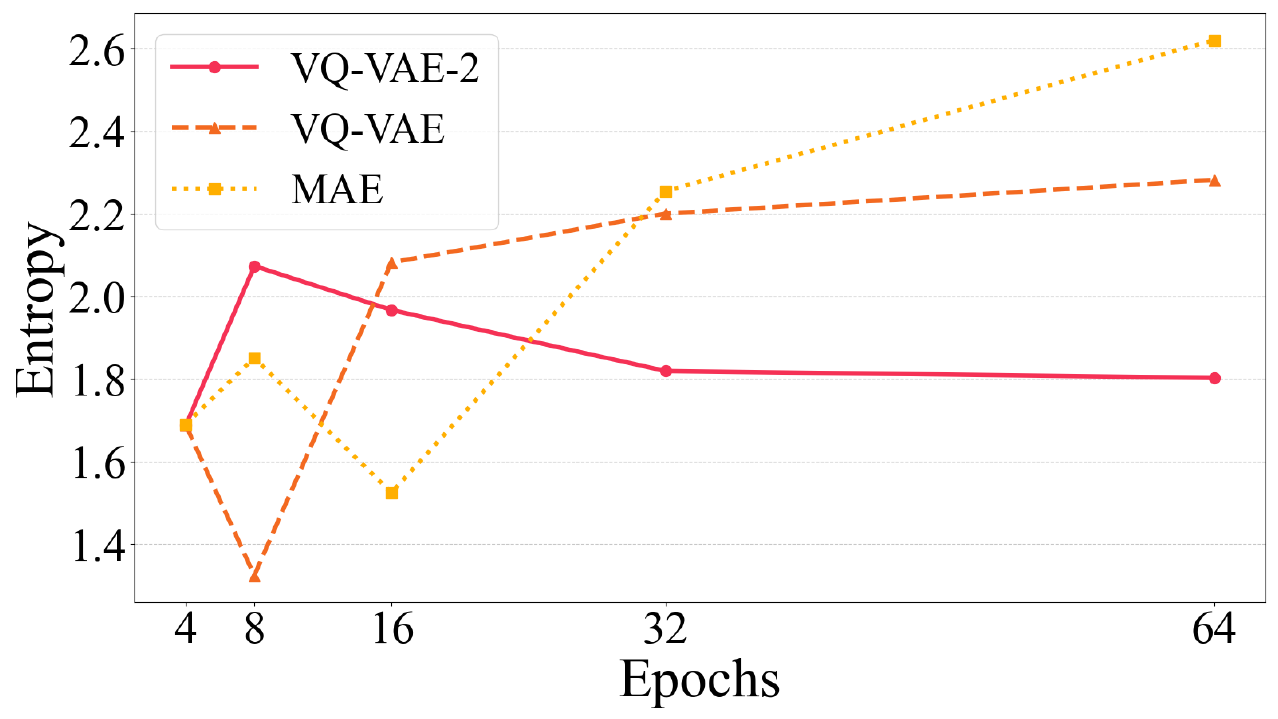}
    \caption{
    The entropy-based quantification of degradation diversity in each dataset. The horizontal axis represents the number of training epochs for each model, and the vertical axis denotes entropy. The dataset generated by VQ-VAE-2 at epoch 8 exhibited the highest entropy among those generated by the same model, indicating a diverse range of degradations. In contrast, datasets generated by VQ-VAE and MAE at epochs 32 and 64 exceeded this entropy value. However, these datasets did not demonstrate effectiveness.
    }
    \label{fig:entropy}
\end{figure}

Figure \ref{fig:entropy} shows the entropy of each dataset. The horizontal axis represents the number of training epochs, and the vertical axis indicates the score. The dataset generated by VQ-VAE-2 at epoch 8 exhibited the highest entropy among the VQ-VAE-2-generated datasets, indicating a high diversity of degradations.  The model fine-tuned with this dataset achieves the highest SR performance, suggesting that degradation diversity contributes to performance improvement.
However, despite the high entropy of the datasets from VQ-VAE and MAE at epochs 32 and 64, it failed to improve the performance of SR models. This suggests that these datasets may contain degradations that negatively impact SR performance.

\subsubsection{Color Difference}
    From Section \ref{sec:deg}, the datasets from VQ-VAE and MAE contain diverse degradations. However, these datasets did not improve SR performance on real-world LR images. A previous study~\cite{FSSRGAN} reported that color differences between HR and LR images can degrade training performance. Based on this finding, we hypothesize that VQ-VAE- and MAE-generated datasets contain significant color differences, which may have caused the performance degradation.
    To validate this hypothesis, we conduct two experiments. First, we quantitatively evaluate the color differences in datasets from VQ-VAE, VQ-VAE-2, and MAE, testing the assumption that these datasets exhibit larger color differences. Second, we assess SR models trained on datasets with color differences, demonstrating that such differences negatively impact training. \\

\noindent \textbf{Pre-processing and evaluation metric.} 
Since color difference cannot be calculated when the sizes of two images are different, the HR image is downsampled using the bicubic method to match the size of the LR image. The color difference caused by downsampling was considered negligible.
The color difference is calculated by using the $\texttt{color.deltaE\_ciede2000}$ module from the Python library $\texttt{skimage}$, adopting the CIEDE2000 metric in the LAB color space~\cite{CIE}. \\

\begin{figure}[!t]
    \centering
    \includegraphics[width=1\linewidth]{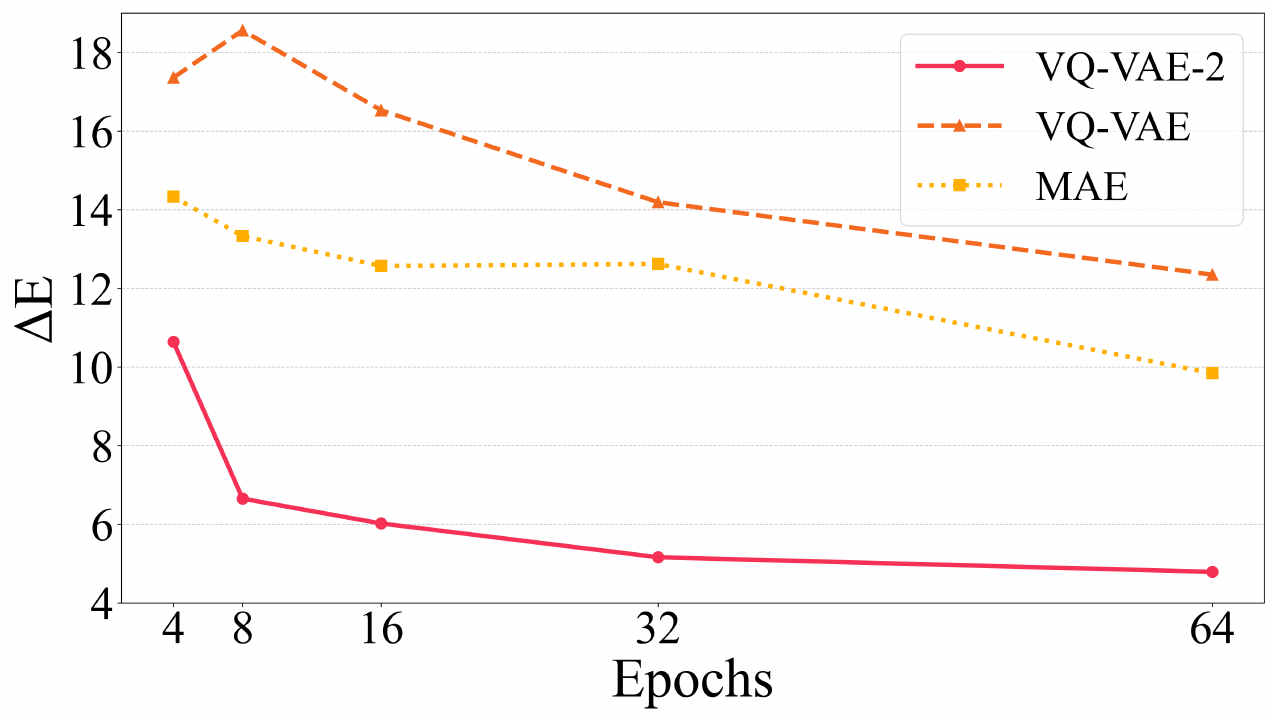}
    \caption{
    Color differences in datasets from VQ-VAE, VQ-VAE-2, and MAE. The horizontal axis represents the number of training epochs, and the vertical axis denotes the color difference $\mathnormal{\Delta} \mathrm{E}$ between HR and LR images. The datasets from VQ-VAE and MAE had large color differences from the early training stages, and even after 64 epochs, the differences remained large. In contrast, the dataset from VQ-VAE-2 exhibited smaller color differences from the early stages and ultimately reduced them to less than half. 
    }
    \label{fig:ep_deltae}
\end{figure}

\noindent \textbf{Color difference analysis of the generated datasets.} 
The comparison of color differences across datasets generated by three models revealed that the datasets from VQ-VAE and MAE exhibit larger color differences than those from VQ-VAE-2.  
Figure \ref{fig:ep_deltae} shows the experimental results, where the horizontal axis represents the number of training epochs, and the vertical axis indicates the color difference $\Delta \mathrm{E}$ between HR and LR images. 
    The results indicate that the datasets generated by VQ-VAE and MAE exhibited significant color differences from the early training stages, these differences remained even reach 64 epochs. In contrast, the dataset from VQ-VAE-2 exhibited smaller color differences from the early stages and ultimately reduced them to less than half. \\

\begin{figure}[!t]
    \centering
    \includegraphics[width=1.0\linewidth]{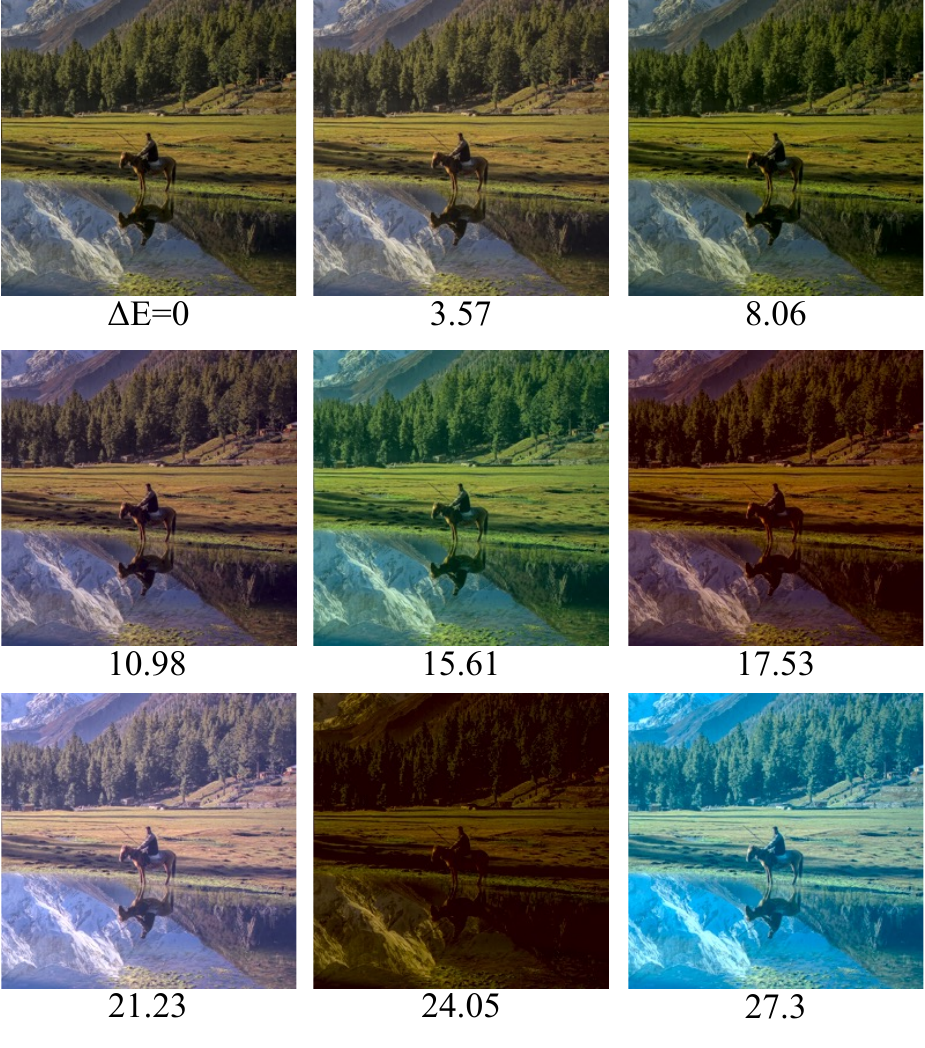}
    \caption{
    The example of LR images from nine datasets used to examine the impact of color differences on training performance. The top-left image is an LR image generated using bicubic interpolation, which introduces no color shift. In contrast, the other images are LR images with color shifts applied in the LAB color space. The values at the bottom of each image represent the average color difference across the entire dataset.
    }
    \label{fig:colordiff_9LR}
\end{figure}

\begin{figure*}[!t]
    \centering
    \includegraphics[width=1.0\linewidth]{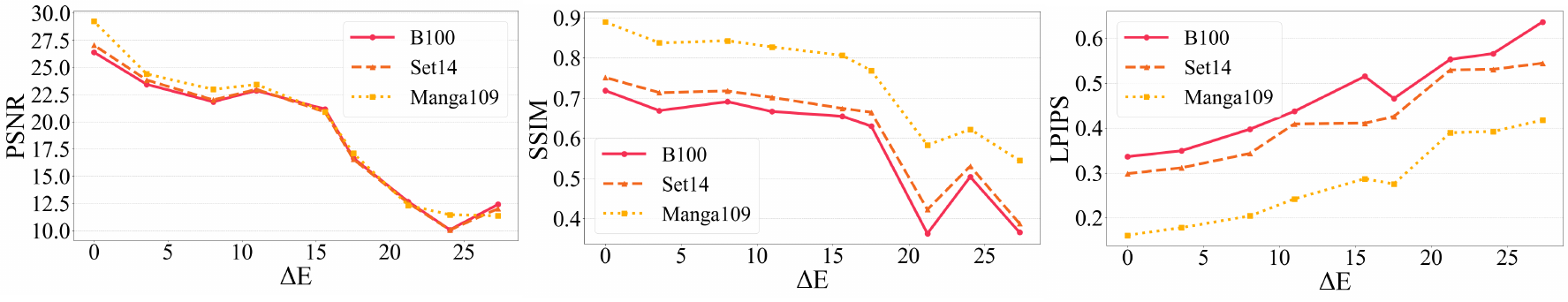}
    \caption{
    The result of training SwinIR on nine datasets with varying color differences. The horizontal axis represents the average color difference of each dataset, and the vertical axis indicates the evaluation metric. The evaluation metric deteriorates as the color difference increases, suggesting that color difference negatively impacts SR performance.
    }
    \label{fig:metrics_deltae}
\end{figure*}

\noindent \textbf{Impact of color differences on training.} 
We next demonstrate that color differences negatively impact the training of SR models. 
In the experiments, we train the SwinIR~\cite{SwinIR} model from scratch for 500,000 iterations using nine datasets. One dataset consists of HR images paired with LR images generated via bicubic downsampling, which have no color difference. The remaining eight datasets exhibit color differences, which were introduced by shifting the colors of the LR images in the LAB color space.
Figure~\ref{fig:colordiff_9LR} presents examples of LR images from each dataset. The top-left image is the bicubic-downsampled LR image without color differences, while the others contain color shifts. The $\Delta \mathrm{E}$ values displayed at the bottom indicate the average color difference across each dataset. DF2K~\cite{DF2K} was used as the HR dataset, and the models were evaluated using three benchmark datasets: Set14~\cite{Set14}, B100~\cite{B100}, and Manga109~\cite{Manga109}.
Figure~\ref{fig:metrics_deltae} shows the evaluation results of SwinIR models trained on these datasets.  The horizontal axis represents the average color difference across the dataset, and the vertical axis represents the metric values. The results show that all evaluation metric scores deteriorate as the color difference increases.
These findings confirm that color differences between HR and LR images in SR datasets negatively impact training. Based on these results, the datasets from VQ-VAE and MAE exhibit large color differences, which likely contribute to their lower training performance. \\

The analysis demonstrates that diverse degradations in datasets can enhance SR performance on real-world LR images. However, among these degradations, color differences may negatively affect training performance and should be excluded from the dataset.

\section{Conclusion}\label{sec:conclusion}
This study proposes a dataset generation method utilizing undertrained image reconstruction models. This approach allows for the generation of LR images with diverse degradations using only HR images. Furthermore, fine-tuning existing SR models using datasets generated by VQ-VAE-2~\cite{VQVAE2}, particularly those trained for 8 epochs, improves various evaluation metrics, reduces blur, and removes noise.
Additionally, an analysis of the generated datasets reveals that incorporating diverse degradations contributes to improving SR performance on real-world LR images. However, among these degradations, color differences between HR and LR images may negatively impact training performance. Future work will focus on analyzing the mechanisms by which VQ-VAE-2 generates effective datasets.

\paragraph{Acknowledgement.}
This work was supported by JSPS KAKENHI Grant Number JP23K24914 and JP22K17962, Japan.

\bibliographystyle{IEEEtran} 

\end{document}